\documentclass{article}

\usepackage{arxiv}

\usepackage[utf8]{inputenc} 
\usepackage[T1]{fontenc}    
\usepackage{hyperref}       
\usepackage{url}            
\usepackage{booktabs}       
\usepackage{amsfonts}       
\usepackage{nicefrac}       
\usepackage{microtype}      
\usepackage{lipsum}		
\usepackage{graphicx}
\graphicspath{ {./figures/} }
\usepackage{natbib}
\usepackage{doi}

\usepackage{amssymb}
\usepackage{amsthm}
\usepackage{amsmath}
\usepackage{amsfonts}
\usepackage{bm}
\usepackage{xcolor}
\usepackage{comment}
\usepackage[export]{adjustbox}
\usepackage{appendix}

\newcommand{\beq}{\begin{equation}}
\newcommand{\eeq}{\end{equation}}
\newcommand{\ba}{\begin{eqnarray}}
\newcommand{\ea}{\end{eqnarray}}

\newcommand{\R}{I\kern-0.3emR}

\title{{\it Homo Entropicus}, the emotional agent and societies of Neural Networks}

\author{Felipp Alves\\
        \texttt{falves@if.usp.br} 
        \AND
        Nestor Caticha\\
        \texttt{ncaticha@usp.br}\\
        Instituto de Fisica\\Universidade de Sao Paulo\\
        CP 66318, 
        CEP 05315-970, S\~ao Paulo, SP, Brazil
        }

\date{\today}

\begin{document}

\maketitle

\begin{abstract}
A neural network with a learning algorithm optimized by information theory entropic dynamics is used to build an agent dubbed {\it Homo Entropicus}. The algorithm can be described at a macroscopic level in terms of aggregate variables interpretable as quantitative markers of proto-emotions.  
We use systems of such interacting neural networks to construct a framework for modeling societies that show complex emergent behavior. 
A few applications are presented  to investigate the role the interactions of opinions about multidimensional issues and  trust on the information source play on the state of the agent society. These include the case of a class of $N$ agents learning from a fixed teacher; two dynamical agents; panels of three agents  modeling the interactions that occur in decisions of the US Court of Appeals, where we quantify how politically biased are the agents, how trustful of other agents-judges of other parties, how much the agents  follow a common understanding of the law. Finally we address under which conditions ideological polarization follows or precedes affective polarization in large societies and how simpler versions of the learning algorithm may change these relations. 
\end{abstract}


\section{Introduction}
\cite{jagerEROS2017}, inspired by Rosaria Conte, has made the case against keeping agents models too simple and ``enhancing the realism of simulations'' (EROS). Realism means psychological realism.  Here we try to address the unneeded gap brought by extreme simplicity. A possible strategy could consider constructing models of agents  which behave according to a minimal list of psychological traits. There is a need to include in the agent's model some complexity that at least remotely resembles some crude emotions but not being able to quantify emotions {\it a priori}, we have to follow a different path. We don't put emotions by hand, but they emerge as natural ingredients in the learning algorithms deduced from optimizing information processing capabilities of agents. 

The agents interact by exchanging opinions represented by binary for/against options or by a continuous variable, about issues that are multidimensional.  A natural choice is to let the agent be an adaptive neural network. The information exchange elicits learning and we describe  optimized learning algorithms deduced from considerations of entropic dynamics in information theory. While mathematically intricate when described by microscopic variables living in possibly  high dimensional spaces, there are simple aggregate or macroscopic variables that estimate what  can be described in terms of words such as surprise, distrust of other agents, confidence in the opinions, performance monitoring. These agents are  capable of learning or reversal of learning. Estimating these quantitative proto-emotions markers,  which emerge from impositions of better   information processing, requires extra machinery (calculations) which can be thought as modules added in an evolutionary selection sequence that enhances information processing obtained by simpler algorithms. By damaging them and measuring performance in learning a fixed rule we can see how the modules influence each other, suggesting evolutionary paths towards the complexification of the information processing agent.

{\it Homo entropicus}, the agent formed by a neural network architecture and the learning algorithm,  is constructed in order to optimally learn a fixed rule from information received from another agent, the emitter, in the form of examples pairs: an issue, parsed into an  input array and an opinion about it. Thus its optimization occurs in the case of small number of agents, in parallel to our evolution in  small societies in the savannas of Africa. This course of action is also related to our incapability to do the necessary mathematics to optimize in larger groups. Again, similarly to humans, we also will take {\it Homo entropicus} from the entropical savannas, its optimizing environment, and ask it to engage in interaction with large groups of similar agents.  

The mathematical deduction of the agent's learning algorithm, which is our main contribution,  is presented very briefly, but we extensively discuss its interpretation in terms of macroscopic aggregate variables and their proposed role as proto-emotions quantifiers. 

The interacting agents framework we have constructed is then applied to different situations, which we now describe.
In order to warm up, we study a class of $N$ students learning form a fixed (non-dynamical) teacher. The dynamical evolution of distrust of the agents towards the static teacher, depends on initial values of the combined effect of previous  knowledge about the subject matter and initial distrust towards the teacher. We don't identify the nature, e.g.  race, gender or dislike of subject matter, of  the initial distrust. Here we start to see how distrust of the source leads to reversal learning and eventual negation of the source's message.  

Next we analyze the case  of the US Courts of appeals in order to inquire as did  \cite{Sunstein2006} whether judges act politically. This is an interesting question since judges are appointed by elected politicians that belong to either one of two parties. They have collected and analyzed data for the US federal
judiciary, where panels of three judges interact and  emit binary opinions about several types of cases. 
We simulate  panels of three interacting agents and measure their opinions about synthetic cases. Different behaviors emerge as we consider the initial conditions, not only of distrust and their confidence about it, but about the initial knowledge about the law and about their political ideology. As discussed in \cite{Sunstein2006}, judges turn out to act as if politically motivated, but their political bias, in addition to their own political party, depends on the other 2 judges in the panel. We find  under which conditions  our agents behave in a statistically analogous form.   If our agents are to replicate the data, they have not only to  trust judges of their same party, but need to extend this courtesy to judges of the other party. In addition, they have to be quite confident about their trust on judges on the opposite camp. For agents that trust and are certain about their trust, we can analyze the behavior when they don't follow a common law, but were only politically motivated and also when they only follow the law without any political bias. We conclude that there is no evidence of those behaviors that can be gleaned from the data.

Polarization in societies, which is the last topic investigated, can come in different flavors as pointed out in \cite{KlarAffective}\cite{IyengarAffective}\cite{Jost2009a}. In addition to empirical measurement, theoretical analysis of their cross influences is under intense debate. What are the mechanisms of interaction between affective and ideological polarization? Do humans dislike each other because they disagree on the issues or is it ideological differences that precede and induce dislike?    We show that the complexity of the agenda under discussion has an influence on the order in which the polarizations occur. When many issues are being discussed ideological polarization precedes affective polarization, but for smaller agendas fast affective polarization is established before ideological polarization. The richness of the agenda also can induce long transients with glassy dynamics and persistent breakdown of balanced norms of trust \cite{Heider}  \cite{HarariBalance} in triads of agents,  as well as breakdown of balance of ideological alignment, which will be fully discussed elsewhere. Among other interesting topics that we will study elsewhere using our framework, we mention societies where some agents  feel disgust or political disdain towards those that vocally express their political opinions, independently of their ideological alignment; agents with for, against or don't know opinions;  the crystallization of discrimination, the effect of different initial conditions of education on inequalities and the study of dictators in these toy societies.

\section{Proto-emotions in neural networks}
We could use a neural network with any architecture with a binary for/against, $\pm 1$ output, however while we want some elements of EROS, this should be tempered with a little KISS and thus we choose as a model for each agent the neural network of simplest architecture, the single layer perceptron with an optimized algorithm. Linearly separable models, in some manner similar to the Rescorla-Wagner model \cite{RescorlaWagner} have been shown to be useful in describing human performance in several cases.

It is natural to consider online learning since opinions about issues are not considered in batches, but only one at a time. The deduction of the general algorithm  can be found in \cite{CatichaEdnna2020}. These methods can be applied, in principle, to more complex architectures, but it is sensible to start with this already rich case since, in addition to being analytically tractable, it offers the possibility of discussing it with high level aggregate variables interpretable as proto-emotions.

Questions that need to be addressed are: (a) What is it that we mean by a proto-emotion? We will answer this retrospectively, once we have an algorithm to analyze. (b) Can a neural network learn without proto-emotions? The answer is yes, very simple algorithms do indeed work, but (c) is the learning efficient? No. This is the reason proto-emotions arise in neural networks. Optimization of an algorithm depends on the context in which the NN is going to perform and thus different algorithms  may result. For a given context, some variables, which will be associated to a proto-emotion, have to be estimated in order to determine the learning step size and direction of change in an optimal way. They inform modulation systems that set the correct scale of the changes.

The society evolves under a discrete time dynamics. At a given time, two agents are selected to interact and randomly assigned the emitter and receiver roles. From a set, that can be called the agenda, an issue/question is chosen. These are the prominent questions of policy which occupy the public attention. The emitter sends its opinion about the issue to the receiver, which  changes its internal state using the learning algorithm. An agent's internal state includes microscopic variables that determine the ideological opinion about the issue and others that estimate the level of certainty about the ``correctness" or about how certain the agent is about this first set. These are called the variables of the ideological/opinion sector. 
Our agents have evolved in an environment where concealed cheating is a possibility, which means that agents have an advantage if they attribute to other agents a measurement of distrust.    Defining what is trust in a human context is beyond our capabilities, see \cite{YamagishiTrust} \cite{InterdisciplinaryPerspectivesonTrustSchockleyEdit} for a discussion. We just stress that we mean something close to distrust in an individual and not mistrust of institutions. So in addition, there are variables that describe how distrustful the receiver agent is about the other emitter and also how certain it is about that distrust attribution. These are the variables of the affective/distrust sector. From a technical point of view, this means the agents are optimized to receive information through a noisy communication channel and distrust is the theoretical answer to the problem of dealing with corrupted data in a robust way. This topic has a long tradition in this area of inference in general and has been studied in several publications \cite{biehl1995noisy}\cite{CoKiCa96},\cite{copelli1997noise} from the point of view of distrust of the information source. For neural networks agents it was studied by \cite{alves2016sympatric}, \cite{CatAlvEsann2019}.

There are several conceivable ways to choose  the emitter/receiver pair of agents. {Borrowing} language from biology,  \cite{alves2016sympatric} 
use the term {\it allopatric} group formation, when an effective communication barrier is created by repeated disagreement and this pair of agents cease to interact, which permits the dynamical grouping with intra-faction interactions, but not inter-faction interactions.  In allopatric group formation the probability of choosing an emitter depends on the distrust the receiver holds toward it, in addition to any other neighborhood constraints that want  to be imposed. Alternatively, {\it sympatric} group formation is introduced as a process where agents keep interacting despite holding  opposing views on the set of issues. The  deduced dynamics of distrust  permits sympatric group formation through  the possibility of anti-learning, i.e learning  the reverse of the emitter's opinion.

 The optimal 
 learning algorithm in general is discussed in\cite{CatichaEdnna2020}. For opinion and distrust is described below without any deduction details. The interested reader should consult \cite{AlvesCaticha2021A} and \cite{FlipPhd}. 
The output of the receiver agent's neural network classifier at time $t$ is 
\ba 
\sigma_r &=&  \text{sign}(\hat{\bm w} \cdot \bm x_t). 
\ea  
where $\bm x_t$ is the issue under consideration, i.e. an assertion parsed into an array of numbers and $\hat{\bm w}$ are part of the ideological sector variables. 
Both $\bm x_t$ and $\hat{\bm w}$ have the same dimension $K$, which can be chosen as needed for a given application. For example, in modeling Haidt and collaborators \cite{Haidt2001a} Moral Foundations theory with $5$ moral foundations, \cite{CaVi2011} used $K=5$. In this context, each component of $\bm x_t$ represents a quantitative estimate of the moral content of an assertion with respect to a given moral foundation. Each component of the agent ideological state $\hat{\bm w}$ represents the importance of a foundation for that agent. From a Bayesian point of view, $\hat{\bm w}$ is the location parameter of the distribution of probabilities of the weights. This distribution has a covariance matrix $\bm C$, which will represent the uncertainty  about the $\hat{\bm w}$ variables, which will lead to the uncertainty about the opinion about an issue.  For the distrust sector the incomplete information available to the agent is represented by a distribution and thus  two distrust/affective  variables are introduced, the expected value $\mu_{e|r}$, that takes value on the reals, and its standard deviation, $V_{e|r}$. We make the convention that negatives values of distrust means that the emitter is trusted (liked) by the receiver. 
It is significant  that  scalar measures of uncertainty for the opinion and the affinity sectors $\gamma_C$ and $\gamma_V$ respectively, can be defined and calculated. While technically important, semantically  the details are not:
\ba
\gamma_C &=& \sqrt{1+{\bm x}\cdot \bm C_t {\bm x}}, \,\,\,\,\,\,
\gamma_V 
= \sqrt{1+V_{e|r}}.  \label{scales}
\ea 
These set the scale in a convenient manner so that the sectors can be compared during a receiver agent interaction. To do so, we need the ideological and affective fields for the receiver agent, interacting with the emitter that sends information $\sigma_e$ about issue $\bm x$, defined by:
\ba
h_{\bm w}&=&\frac{\hat{\bm w}\cdot\bm x\sigma_e}{\gamma_C}, \,\,\,\,\, h_\mu = \frac{\mu_{e|r}}{\gamma_V},
\label{stabilities}
\ea 
so that $h_{\bm w}>0$ the emitter and the receiver agree on the issue. If $|h_{\bm w}|$ is large, the receiver is very sure about its opinion. If $h_{\mu}>0$ the receiver distrusts the emitter and if $|h_{\mu}|$ is large, it is very confident about the distrust. Given these preliminaries, the interaction dynamics, leads to the new state of the receiver: 
\ba
    \hat {\bm w}_{t+1} & =& 
                   \hat {\bm w}_{t} + \frac{1}{\gamma_C}F_w {\bm C_{t}{\bm x}\sigma_e} \label{eq:edw}\\
    \bm C_{t+1} & =&
              \bm C_{t} + \frac{1}{\gamma_C^2}F_C \bm C_{t}{\bm x}{\bm x}^T\bm C_{t} \label{eq:edc}\\
    \mu_{e|r}(t+1) & = &
                 \mu_{e|r}(t) + \frac{1}{\gamma_V} F_m {V_{e|r}(t)}\label{eq:edm}\\
    V_{e|r}(t+1) & =&
                 V_{e|r}(t) + \frac{1}{\gamma_V^2}F_V {V_{e|r}(t)^2}\label{eq:edv}
\ea
Equation \ref{eq:edw} describes the changes of the ideological sector variables. 
First note the term ${\bm x}\sigma_e$, this is called the Hebbian term in the Statistical Physics community of neural networks. If we forget the prefactor,  it just adds to $\hat {\bm w}_{t}$ the issue vector $\bm x$ if the emitter's opinion is positive or the reversed sign issue $-\bm x$, if the emitter's opinion is negative. This is the simplest learning algorithm, which when applied to learning a fixed rule, leads to a decay of the probability of disagreeing with the emitter, i.e. the generalization error,  asymptotically proportional to $1/\sqrt{P}$ after learning $P$ independent issues. This is an algorithm that treats every issue in the same manner, there are no emotions. With the full algorithm the behavior saturates theoretical bounds, with an asymptotic generalization error decaying as $1/P$. 
So, let's not forget the whole prefactor. The Hebbian term is changed to ${\bm C_{t}{\bm x}\sigma_e} $. This is called a tensorial update. The covariance matrix rotates and stretches the issue $\bm x$ in a non intuitive way. The covariance is changed, as described in  \ref{eq:edc}, by the the addition of a rank 1 matrix, formed by the vector modifying $\hat{ \bm w}_t $.  These elements give a geometrical recipe to the update of the ideological sector variables. 

Given this geometrical structure, the magnitude of the changes is modulated by the functions $F_w$ and $F_c$ which we discuss in detail. Since 
equations \ref{eq:edm} and \ref{eq:edv}, which  give the update of the mean and standard deviation of the affinity sector, also show modulations functions $F_m$  and $F_V$, we discuss them together.  They are functions of the the ideological $h_{\bm w}$ and affective $h_\mu$ fields given by equation \ref{stabilities}:

\ba
    F_w (h_w,h_\mu)& =&  (1-2\Phi(h_\mu))\frac{g(h_w)}{Z},\label{eq:Fw}\\
    F_C (h_w,h_\mu)& = & -F_w(F_w + h_w),\label{eq:FC}\\
    F_\mu (h_w,h_\mu)& =&  (1-2\Phi(h_w))\frac{g(h_\mu)}{Z},\label{eq:Fm}\\
    F_V (h_w,h_\mu)& =& -F_\mu(F_\mu + h_\mu),\label{eq:FV}
\ea
where $g(u) =(2\pi)^{-1/2}\mathrm{e}^{-\frac{1}{2}u^2}$ is a standard Gaussian. $Z$, called the Bayesian evidence for the model of the receiver, is
\ba
Z & = &      \Phi(h_{\bm w}) + \Phi(h_{\mu_{e|r}})  - 2\Phi(h_{\bm w})\Phi(h_{\mu_{e|r}})
\label{eq:evidence}
\ea
and $\Phi(s) = \int_{-\infty}^s\mathrm{d}u g(u)$ is the cumulative distribution of the Gaussian. Equations \ref{eq:Fw} to \ref{eq:FV} can be used to see a symmetry between the two sectors, but only when appropriately scaled by the $\gamma$'s 
\ba
F_w(h_w,h_\mu) &=& F_\mu(h_\mu,h_w), \\
F_C(h_w,h_\mu) &=& F_V(h_\mu,h_w).
\ea 
These expressions, deduced in \cite{AlvesCaticha2021A}, 
are quite interesting and reasonable as we show next. For fixed distrust, the opinion sector has the modulation function obtained for learning optimally in the Tree Committee and the Parity machine neural networks \cite{CoCa1995} \cite{SimCa96} obtained by different mathematical arguments.   While remarkable, this is {\it a posteriori} not surprising, in view of the mathematical equivalence of noise and hidden units in some machines \cite{CoKiCa96}. 
\section{The modulation functions: What should be blamed for a surprise?}

\begin{figure}[h!]
    
    \includegraphics[scale=0.62]{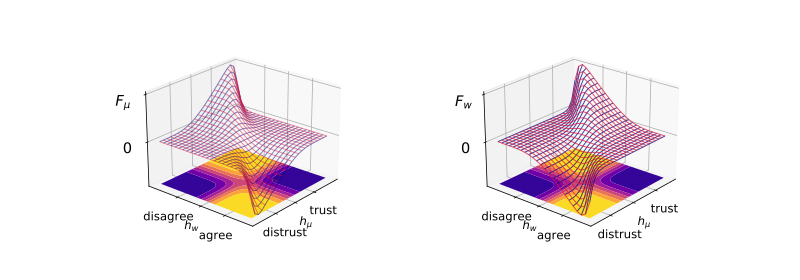}
    \caption{The modulation functions for the affective sector (left) and the opinion sector (right) in the space of agreement field $h_w$ and distrust field $h_\mu$. In the floor the contour plot of the evidence $Z$. Dark regions correspond to corroboration and light regions to surprises. Note that intense learning only occurs in the surprise regions and essentially only in the sector whose field has the smallest absolute value and it is thus blamed for the surprise. Negative values for $F_w$ mean reversal of learning.}
    \label{fig:funcoesmodulacao}
\end{figure}
\begin{figure}[h!]
    \centering
    
    \includegraphics[scale=0.62]{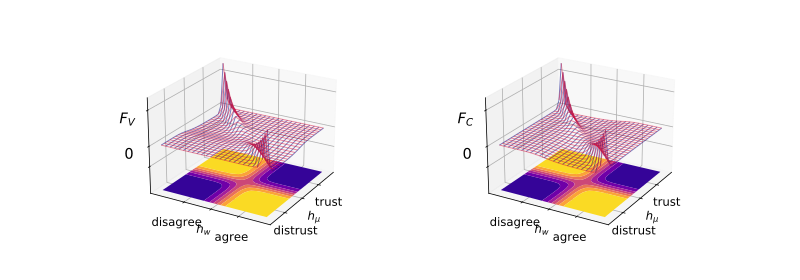}
    \caption{Similar to figure \ref{fig:funcoesmodulacao}, but for the uncertainty modulation functions. The uncertainty increases near the diagonal where $h_w \approx -h_\mu$. A surprise where the other sector is blamed results in a slight decrease in uncertainty.}
    \label{fig:funcoesmodulacaoVar}
\end{figure}
\begin{figure}[h!]
    \centering
    \includegraphics[scale=0.5]{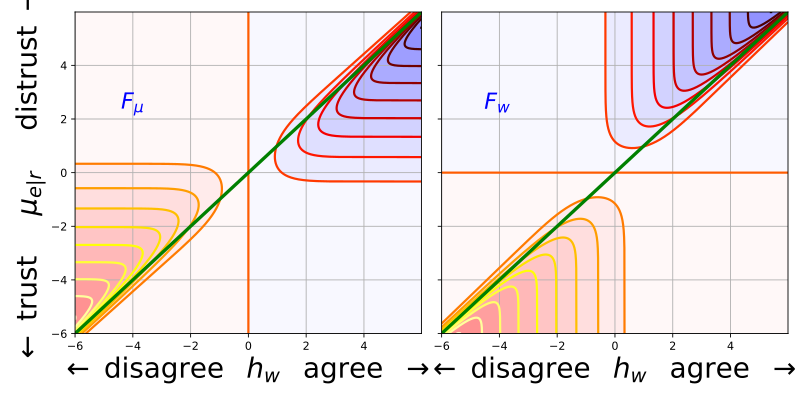}
    \caption{Similar to figure \ref{fig:funcoesmodulacao}. The modulation functions are positive (respect. negative) in the red (blue) regions.}
    \label{fig:funcoesmodulacaoPlana}
\end{figure}
\begin{figure}[h!]
    \centering
    \includegraphics[scale=0.5]{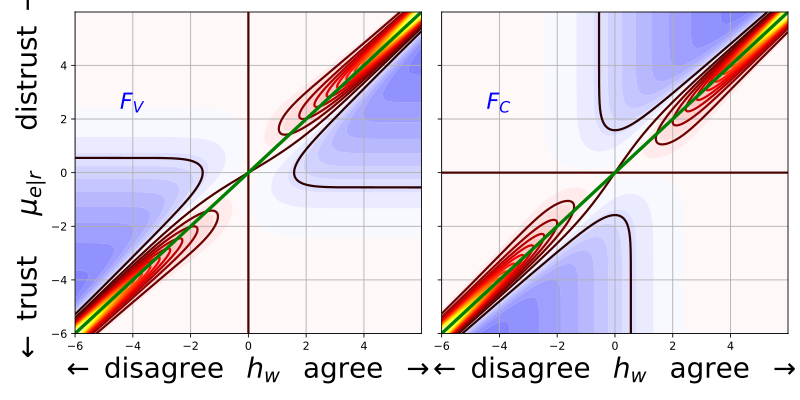}
    \caption{Blame attribution can be seen for the uncertainty modulation functions shown in figure \ref{fig:funcoesmodulacaoVar}. The green line diagonal is the border of blame. The uncertainty increases near the diagonal where $h_w \approx h_\mu$. A surprise, when a  sector is blamed, results in a slight decrease in uncertainty, shown in the light blue regions, since there is effective learning. Near the blame cross-over no update in the uncertainty occurs, although both sectors change their weights. After the blame cross-over, there is a large increase, near the diagonal, in the red-yellow regions.}
    \label{fig:funcoesmodulacaoVarPlana}
\end{figure}
\begin{figure}[h!]
    \centering
    \includegraphics[scale=0.5]{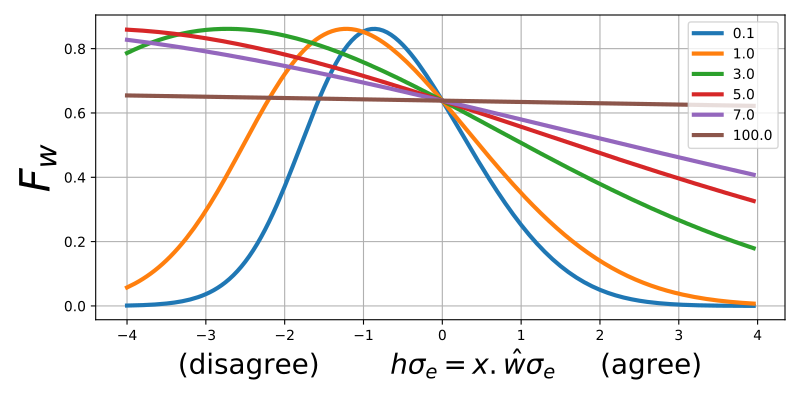}
    \caption{The modulation function $F_w$ as a function of $h= \sigma \bm x\hat{\bm w}$ depends on the agent's uncertainty about its opinion. Very uncertain agents don't modulate significantly different agreements from disagreements. Small uncertainty leads to surprise detection with a protection against learning from extreme disagreement, since it is then attributed to the noisy channel.}
    \label{fig:funcoesmodulacaoVariacional}
\end{figure}
As shown in figures  \ref{fig:funcoesmodulacao}, \ref{fig:funcoesmodulacaoVar} and \ref{fig:funcoesmodulacaoPlana}, \ref{fig:funcoesmodulacaoVarPlana}, the modulation functions for the two sectors are only large in the surprise regions where $h_w h_\mu >0$, the receiver disagrees with a trusted emitter ($h_\mu < 0, h_w<0$) or agrees with a distrusted emitter ($h_\mu > 0, h_w >0$). But this doesn't mean that both sectors will be strongly modulated. There is a crossover region near the diagonal $h_w= h_\mu$, with a transition from blaming  the opinion sector for the surprise, when $|h_w|< |h_\mu|$,  to blaming the affective sector when $|h_w|> |h_\mu|$. We stress that the crossover is clearly seen only  when the fields are correctly scaled (equation \ref{stabilities}) so that their values can be compared. At the diagonal, the crossover of blame attribution leads to a strong signal increasing the uncertainties, not knowing which sector to blame for the surprise boost uncertainty about opinions and mistrust.  

The scaling of equation \ref{stabilities}, by describing the two sectors in comparable units, permits seeing this transition, which appears as very natural, at least in a non quantitative manner.  However it is quite instructive to redraw the modulation function $\gamma_C F_w$ as a function of the raw stability field $h\sigma_e= \bm x \cdot \hat{\bm w}_r \sigma_e$, for different values of the uncertainty scale in the opinion sector, as in figure \ref{fig:funcoesmodulacaoVariacional}, which is known from the student-teacher scenario in Statistical Mechanics \cite{KiCa1993}. In this case the student-receiver trusts the teacher-emitter with $\mu = -3$. Since the answer of the receiver is $\sigma_r= $sign$(h)$, $h$ measures how certain the student is in its opinion about the issue in the sense of how robust is a classification under small perturbations of the issue or of its weight vector. We can now see that for large uncertainties, the student is quite indifferent to whether it agrees or not with the teacher: the modulation under strong uncertainty is oblivious to the magnitude of the internal $h$ and the students treats information independently of its answer. If the student is very uncertain about the rule, an error is not surprising. As learning of the teacher's rule proceeds and the uncertainty decreases, concurring examples will elicit smaller changes, but disagreement ($h<0$) generates a larger scale  modulation. This surprise persists until there is the cross-over to blaming the teacher for the error and the student remains unchanged. For these agents a surprise only occurs if there is some certainty about a prediction. Technically the neural network has gone from learning by correlations in the initial stages to learning by strong error correction, modulated by a measure of the surprise, with a protection against a drunk teacher. This optimal rule learning scheme depends not only on the Hebbian correlation term ($\bm x \sigma_e$), but on extra machinery that can be though of as added hardware, by hand or by evolution, if these machines were under pressure to generalize better. One new module to measure surprises and hence correct errors, another to estimate performance in order to estimate uncertainty. It is interesting that measuring only surprise and not performance leads to a better generalization than the pure Hebbian, however measuring only performance and not surprise is equivalent to Hebbian generalization. Were this an evolutionary process and a cost be attached to the acquisition of these pieces of hardware, the surprise measuring module would be selected before the performance hardware. There is no advantage to be gained by using hardware that doesn't  improve the generalization. Alternatively, a surprise module alone does provide generalization improvement and so will invade a population. This is further discussed in \cite{CaKi97TimeB} and \cite{Neirotti2003}. This is consistent, at least at a descriptive level, with fact that the amygdala, which shows intense activity when associated to surprise and fear, appeared before frontal lobe, involved in performance evaluation, in the evolution of vertebrates.

\section{Societies of Neural Networks}
See \cite{Metzler00} for early work on interacting perceptrons. Our model is already sufficiently complex and that some empirical facts can be understood doing this intermediate step justifies that we don't look directly at richer agent models. Nevertheless, our claim is that the natural way to construct realistic agents that decide and influence others is based on information processing machines, which find  their current paradigm in neural networks. In the following examples we study frist a class of $N$ students learning form a teacher, which remains fixed. The students don't talk to each other. This is not a realistic but, a necessary step to understand the behavior in simple cases.  This is a warm up for a more realistic application, where three agents model a panel of judges. What makes it interesting is the availability of data, collected by \cite{Sunstein2006}. It permits quantification of the political bias and the degree to which judges, collectively follow the law. In the future, this will be extended to the analysis of specific US appeal courts. We work with for US data for two reasons, the availability of the data and the simplicity of analysis brought forward by their simple two party culture. A preliminary report appeared in \cite{CatAlvEsann2019}. 
The last application discusses how the different sectors of our agents lead to introducing two types of polarization and the breakdown of balanced affections and ideologies, contributing to modeling the current discussion on affective versus ideological polarization, studied by the political science community \cite{KlarAffective_Full}\cite{IyengarAffective} by identifying conditions where one or the other polarization is established first. 

\subsection{A class of $N$ students and one teacher \label{oneteacher}}

\begin{figure}[b!]
    \centering
    \includegraphics[scale=0.3]{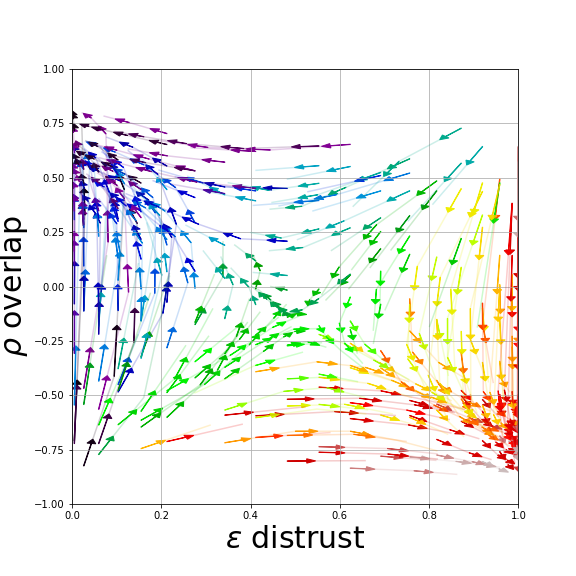}
     \includegraphics[scale=0.3]{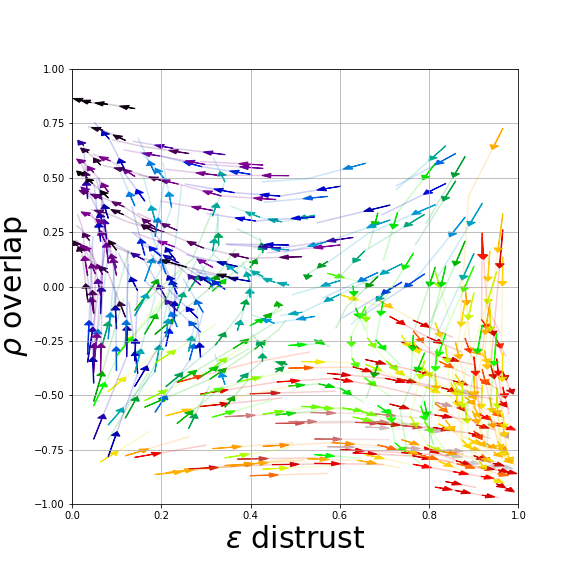}
    \caption{An example of a simulation of class of $N=300$ student agents interacting with a teacher network.  Color coded by ranking of final grade. Color goes from violet to red as overlap of student-teacher decreases from high to low values.
    Trajectories for every student in the $\varepsilon, \rho$ space. Left: $P=1$, right: $P=300$.}
    \label{fig:ClassNagents}
\end{figure}
\begin{figure}[hb!]
    \centering
    \includegraphics[scale=0.5]{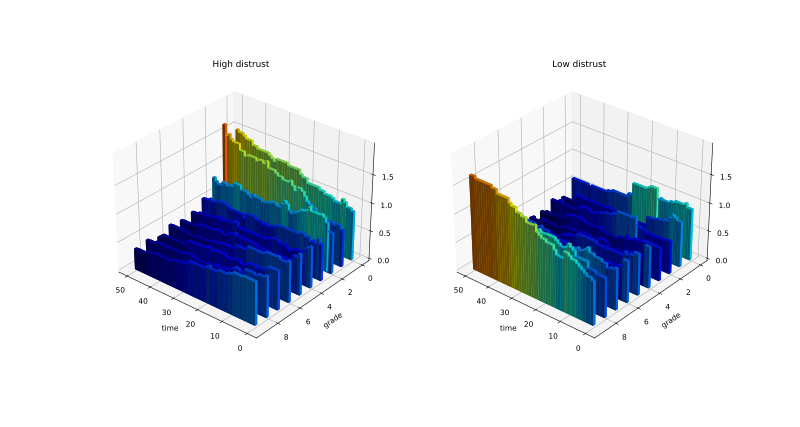}
    \caption{Same case as in left side of figure \ref{fig:ClassNagents}. Time evolution of distribution of grades for initial high distrust ($\varepsilon>1/2$, left), low distrust ($\varepsilon<1/2$, right). As time increases the inequality of the distribution of grades stems from the initial distrust of the teacher.  }
    \label{fig:histogramas}
\end{figure}
In this section we show an example from a student-teacher scenario. At every step of the dynamics one of $N$ students is uniformly chosen independently of anything else and it learns the opinion of a fixed teacher about a vector randomly chosen from the learning set. We see that some  student agents learn and other don't, despite having similar previous knowledge of a subject. In order to build into the  model 
initial conditions that describe the agent we proceed as follows. First all vectors that appear on the right hand side of the expressions below have unit length. The ones that appear at the right are also normalized to one.
The teacher or emitter would like to teach a {\it curricular core} vector $\bm V_c$. Students will be graded along their dynamics based on their similarity to  $\bm V_c$: the overlap $\rho_i=\bm w_i \cdot \bm V_c/|\bm w_i|$. The learning set has $P$  vectors $\bm V_a$. Each $\bm V_a$  is constructed by adding to $\bm V_c$ a neutral contribution $\bm N$ orthogonal to $\bm V_c$ and a random vector $\bm \eta_V$:
\beq
\bm V_a  \propto \bm N + \alpha_v \bm V_c  +\bm \eta_a.
\eeq 
To model diversity in the student's previous knowledge about the subject being taught, their initial weight vectors receives a contribution $ \bm v_\theta$ of a random vector  making an angle $\theta$ with $\bm V_c$. A random vector $\bm \eta_i$ of unit length is added to model peculiarities of each student: 
\beq 
\bm w_i(t=0) \propto \bm v_\theta +\alpha_P\bm \eta_i,
\eeq
where $\alpha_P$ measures the relative importance of such contributions. The teacher is fixed at:
\beq 
\bm w_T \propto \bm \eta +\alpha_T \bm V_c .
\eeq 
The parameters $\alpha_V$ and $\alpha_T$ control the importance of the subject on the issues being discussed and on the teacher respectively. The choice of the initial values of the  set of distrusts $\{\mu_{r|e}\}$ define the initial distrusts
\beq
\varepsilon_{e|r} = \Phi(\mu_{e|r}),
\eeq 
which are chosen iid uniformly in the interval $(0,1)$. The informational interpretation of $\varepsilon$ is that it is the probability of the "correct" answer being flipped in the noisy channel. If $\mu_{e|r}<0 $(respectively $>0$) then $\varepsilon<1/2$ ($>1/2$) and the information source/teacher is initially trusted (distrusted) by the student. 

Results from a simulation in figure \ref{fig:ClassNagents} show typical flows in the space $(\rho,\varepsilon)$, with an unstable hyperbolic fixed point at $\rho=0, \varepsilon=1/2$. The effect of changing the number of issues $P$ is simply to go from quite smooth hyperbolic trajectories for one example,   to a more disordered flux as $P$ is increased to $300$. Regardless of $P$, we can clearly see that the class polarizes into two groups. Students with good grades and small distrust in one group and bad grades and large distrust for the teacher in the other. Interestingly both the certainty about the answers and for the distrust increase in average, students become more sure about their opinions independently of their grades. 
This is of course due to the crossover blame attribution that emerges from the symmetry of the modulation functions. The expected grades histograms as a function of time is shown in figure \ref{fig:histogramas}. The students are separated into to groups according to initial distrust of the teacher and the histograms show the evolution in time of the class. Grades are reported on a 0 ($\rho =-1$) to 10 ($\rho =1 $) scale. 

From the point of view of the interpretation it is interesting that a strong distrust for a teacher can lead an initially capable student to have a terrible final performance. Symmetrically, an initially  very bad student can achieve good grades if the initial distrust is weak. These scenario can also be interpreted as a model for the development of negation despite receiving positive information. We will present elsewhere applications to climate and vaccine negationism,  as well as differences of performance in testing by different ethnic groups. It is clear that we can model priming the students by boosting the initial distrust about the information source in either way.  Variations on the theme, with different parameters and conditions, will be studied elsewhere. 

\section{US court of appeals: shared knowledge of the law and  political bias.}
\begin{figure}[b!]
    \centering
    \includegraphics[scale=0.35]{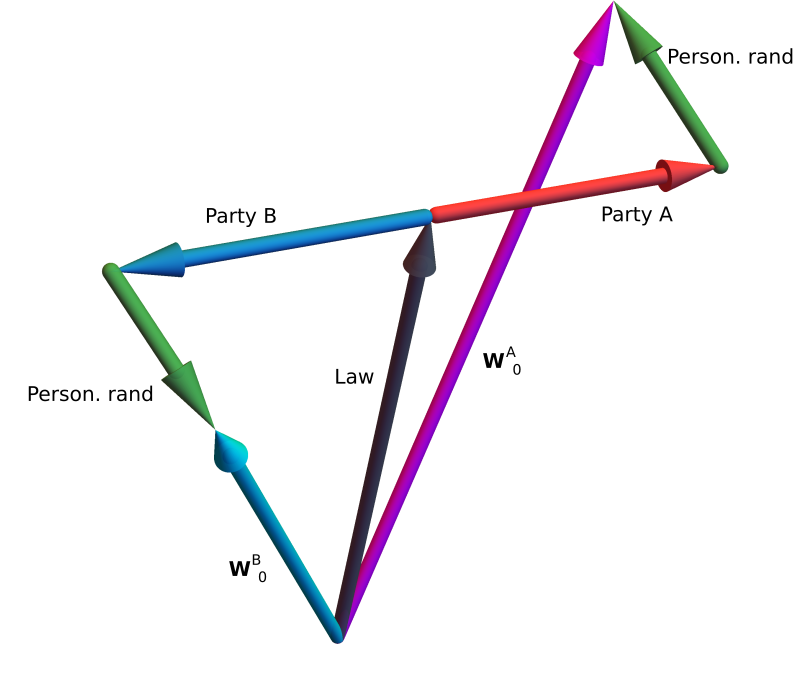}
    \caption{Low dimensional  representation of the initial states of the agents, $\bm w_0^A$ and $\bm w_0^B$ made up by three contributions. A vector $\bm L$ ({\it Law}) is common to all agents. The direction of the {\it party} contribution depends on the party being $A$ or $B$. A random contribution that is unique to each agent models the idiosyncratic part of a judge. 
    }
    \label{fig:initialjudges3agents}
\end{figure}

 Three member panels of USA Courts of Appeals judges presents a good opportunity to apply our model to the three agent case, since besides its intrinsic interest, there is data  \cite{Sunstein2006} kindly shared by A. Sawicki. They  investigated the influence on decisions of political party alliances, i.e the party that appointed them. We construct statistical signatures of behaviors that can be measured in both the theoretical and real systems and simulate the model for different sets of initial conditions scenarios. The scenarios can be manipulated by changing (a)  the magnitude of the influence of the Law that is common to all agents,  (b) the magnitude of the party's influence on the initial state of an agent, (c) the relative contribution of 
its idiosyncratic component, see figure \ref{fig:initialjudges3agents} and (d) the initial attribution of distrust and its uncertainty about judges of the same and of the other party. Preliminary results can be found in \cite{CatAlvEsann2019}.

From  \cite{Sunstein2006} (Figure 2-2. Voting Patterns for Case Types with Both Party
and Panel Effects)
we obtain  the empirical mean percentages of liberal decisions by judges in three member panels, under different conditions of panel composition for each of the 15  areas of cases
\footnote{The data is from rulings in 15 areas of the law: 
Affirmative action, NEPA, 11th Amendment, NLRB, Sex discrimination, ADA, Campaign Finance, Piercing corporate veil, EPA,
Obscenity, Title VII, Desegregation, FCC, Contract Clause, Commercial speech.} ranging from very ideological to not ideological.
Judges, who are appointed by an executive officer, are labeled by the officer's party . When focusing on the decisions of a republican indicated judge in a panel of two republicans and one democrat, we use the notation $v=Rrd$, capitalizing the initial of the party of the judge under observation. There are six different types of votes $v$: $Rrr, Rrd, Rdd, Drr, Drd, Ddd$. 

The data supports their three working hypotheses, that  there is (i) Ideological voting: ``Republican appointees vote very differently from Democratic appointees";
(ii) Ideological dampening: a judge in the minority party of a panel will be less ideological; and (iii) Ideological amplification: a judge in a pure party panel will be more ideological. Hence they are describing the interactions of the judges in the panel. 

We represent their data  as a set of 15 dimensional vectors $\bm{J}_v$, one vector for each $v$. A component
of $\bm{J}_v$ is the percentage above  $50\%$ a judge in a panel $v$ votes in favor of the liberal position for a particular type of case. The angles $\theta(v,v')$ between these vectors measure differences between judges in different panels.  For instance $\theta(Rrr,Ddd)$ measures the difference between Republicans and Democrats in pure panels, hence permitting to assess hypothesis (i) of Ideological voting. Comparing $\theta(Ddd,Rdd)$
and $\theta(Ddd,Drr)$ informs about how liberal is a Republican sitting with two democrats and whether it is more so than a Democrat sitting with two republicans, hence probing hypothesis (ii). 
The angle $\theta(Rdd,Rrr)$ measures the differences of judges in the minority or in a pure panel, relevant for hypothesis (iii). The angles $\theta(Rrr,Rrd)$ or $\theta(Drr,Drd)$ inform about the differences that occur in panels where a companion judge from one party is changed to the other party. 

The introduction of the arrays $\bm{J}_v$ and the angles between them gives a quantitative meaning to the idea of alignment of views.
The main reason to introduce $\bm{J}_v$ is that it can be constructed from readily observable quantities. 
We can define angles between the vectors that represent the state of the agents $\bm w$, but these are not empirically available 
for the judges since $\bm w$ states are only indirectly hinted from voting patterns.

For the  model we consider a two parties ($A$ and $B$) system. 
Three adaptive agents, each representing a judge, interact by exchanging their opinions about a particular issue to be judged. The initial state $\bm w_{i|I}$, where $I=A$ (respect. $I=B$) for agents appointed by party $A$ (respect. by party $B$), of an agent at the beginning of a discussion reflects three
main characteristics the judges ought to have: a common knowledge of the law; shouldn't have: an ideological bias that depends on the political party $I$ of the executive officer that made the appointment; and simply have: an idiosyncratic random characteristic. Then
\beq 
\bm w_{i|I}(t=0) = \alpha_L \bm{L} \pm \alpha_P\bm{P} +\alpha_\eta\bm{\eta}_i,
\eeq 
as illustrated in figure \ref{fig:initialjudges3agents}. 
The first term $\bm L$
represents {\it knowledge} of the Law, common to all agents. If this were the only term, agents would have identical opinions on every issue. The second term $\bm P$ represents ideological party lines, perpendicular to $\bm L$. The plus and minus signs indicates an agent appointed by party $A$ or $B$, respectively. The third term $\bm \eta_i$ is a vector independently chosen uniformly at random on the unit sphere in $K$ dimensions for each agent.
The parameters $\alpha_L, \alpha_P$ and $\alpha_\eta$ control the relative importance of the Law, the party and personality of the agents. The cases $\alpha_L =0$ can be called  {\it Lawless} cases and if $\alpha_P=0$ {\it party-less} cases. 

For the initial distrust attribution we consider four different  scenarios, see table \ref{initialcond}.
By courteous we mean    initially the agents, independently of party, extend the courtesy of attributing a low distrust value of $\mu_{a|a}=\mu_{b|a}=\mu_{a|b}=\mu_{b|b} =-1 $ to each other. Discourteous means that agents initially trust same party agents $\mu_{a|a}=\mu_{b|b} =-1 $ and distrust agents form the other party $\mu_{b|a}=\mu_{a|b}=1 $. A low $V_0 =0.1$ means the agents are quite certain of an attribution of distrust and a high value $V_0= 5$ that agents are uncertain of the attribution.

\begin{table}[htb]
\centering
Distrust Scenarios\\
\begin{tabular}{ | l l | c | l | }
\hline
\hline
 &\multicolumn{1}{c|}{Initial Conditions } &\multicolumn{1}{c|}{Courteous} &\multicolumn{1}{c|}{Discourteous} \\
&\multicolumn{1}{c|}{ scenarios} &\multicolumn{1}{c|}{} &\multicolumn{1}{c|}{} \\
 \hline
 & Certain  &  $\mu_{e|r}=-1$, $V_0 = 0.1$ & $\mu_{e|r} =1,$ $V_0 = 0.1$\\
\hline
 & Uncertain  &  $\mu_{e|r}=-1$, $V_0 = 5.0$ & $\mu_{e|r} =1,$ $V_0 = 5.0$ \\
\hline
\end{tabular}\caption{Four different distrust initial conditions scenarios were considered. $\mu_{e|r}$ refers to agents of different parties. For all pairs of agents of the same party $\mu=-1$ and $V_0=0.1$  }
\label{initialcond}
\end{table}

An issue, characterized by its angle $\phi$ with the Law vector, is chosen and the judges engage in the exchange of opinions.  We repeat this for $n_{case}=15$ different $\phi$ angles, taken uniformly in the interval $[-\pi,\pi]$. This is repeated for a few hundred sets of initial conditions. For each run we record the voting patterns, and the averages are used to construct the
$n_{case}$ dimensional  $\bm{J}_v$ vectors. Then, repeat for all $v$ environments.  
For simplicity, we froze the $C$ dynamics at $C_{ab}=2 \delta_{ab}$ since it has been shown to be related to cognitive style, which we expect to be frozen when the agents enter adulthood \cite{CaVi2011, CaCeVi2015}.

\begin{figure}[ht]
\includegraphics[width=1.\linewidth]{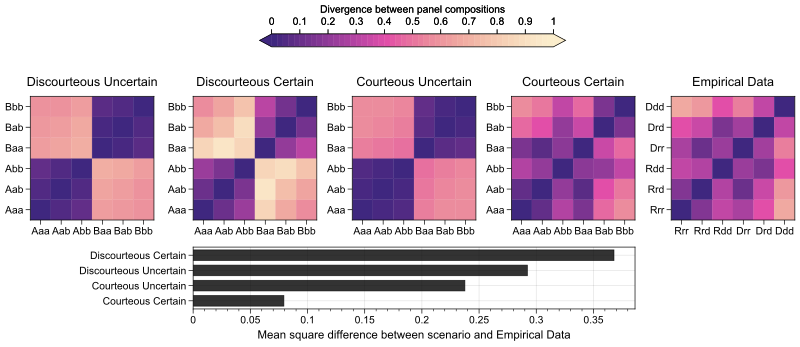}
\caption{Voting alignment as represented by the matrix of angles $\theta_{vv'}/\pi$ between vectors $\bm{J}_v$ and $\bm{J}_{v'}$.  Initial values $\bm \alpha =(1,1,1)$. Top: four different initial conditions for the affective sector lead to the steady state figures, which color code for the angles between agent-judges in the possible panel compositions. Far right: the same representation of the empirical data for human-judges. Bottom: quantitative difference between the different scenarios and the empirical data.  Visually and quantitatively, the courteous/certain scenario are closest to the empirical data.}
\label{bandeiraspanel1juc}
\end{figure}
\begin{figure}[ht]
\begin{center}
\includegraphics[width=.7\linewidth]{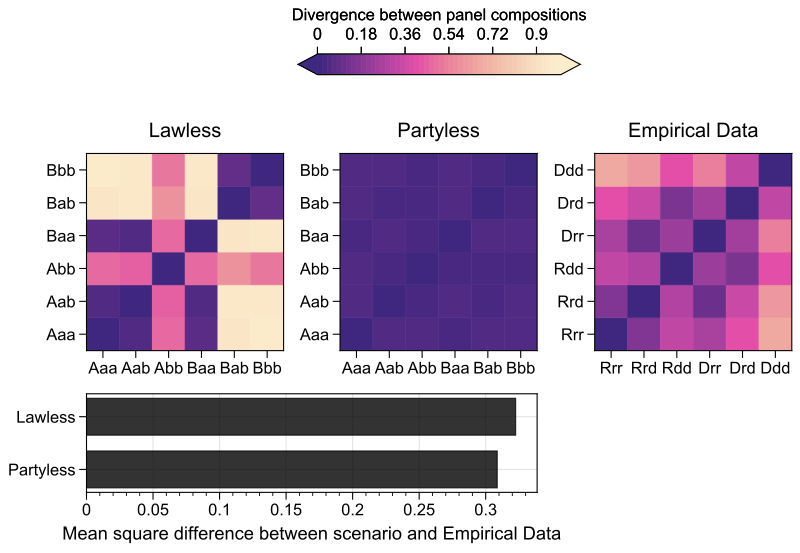}
\end{center}
\caption{Same as figure \ref{bandeiraspanel1juc}. Final states of voting alignment ($\theta_{vv'}/\pi$),  for agent/judges with, at left no initial component of the law $\bm \alpha = (0,1,1)$ and, center, no initial political component,   $\bm \alpha = (1,0,1)$.Right: Empirical values. Neither initial condition leads to anything similar to the empirical data. Sadly, judges are political, but they show some common understanding of the law.}
\label{bandeiraspanel2juc}
\end{figure}
Both from the empirical data and from the simulations we can construct  vectors $\bm{J}_v$ for each of the six types of votes.
The angles between two vectors indicate how two agents (or how two judges) are { \it aligned} in their views. Similar voting patterns will result in small angles. 

In figure \ref{bandeiraspanel1juc}  we present, the angles between the vectors $\bm{J}_v$ obtained from the voting patterns of the agents in the four different trust initial conditions, see Table \ref{initialcond}  and the same statistic for the judges obtained from the empirical data. In this color map representation of the matrix of angles $\theta(v,v')$ between the vectors $\bm{J}_v$ and
$\bm{J}_{v'}$, the light entries represent large angles and different voting patterns, while dark colors mean small angles or very aligned voting patterns. 
For these simulations the original agents weights are generated with $\alpha_L=\alpha_P=\alpha_\eta=1$. In general we see that  party $A$ and party $B$ appointees agents are very different.  It is clear that the courteous-certain scenario ($4^{th}$ from left, figure \ref{bandeiraspanel1juc} ) is qualitatively much closer to the empirical data (far right image). The first three images are obviously visually different from the empirical data. This conclusions are quantitatively supported by the mean quadratic error for over all  panel pairs, shown in the bottom of figure \ref{bandeiraspanel1juc} for the four scenarios. The conclusion is that the model behaves similarly to the Appellate Courts only if the agents trust each other and are quite
certain about this trust at the beginning of the interactions.  

The courteous-certain scenario is summarized in table \ref{tabelaJuiz}. Even in this optimistic scenario,  we see that judges 
appointed by party $A$ behave differently from those appointed by party $B$, hence we see evidence of ideological voting,
a reminiscent behavior of the party dependent initial conditions. 

We also see evidence of Ideological dampening, for example
$\theta(Rrr,Ddd) > \theta(Rrd,Ddd)$ meaning that the difference between a Republican in a pure republican panel
and a Democrat in a pure democratic panel is larger that of the same Democrat and a Republican who is interacting with one republican and one democrat. Interestingly a Republican in the presence of two democrats acts as more liberal that a Democrat in the presence of two republicans. The same results hold if we change R (and D) for A (and B) in the courteous-certain scenario. 
Also  Democrats in the company of one republican and another democrat are  more similar to Republicans in the presence of two 
democrats than to Democrats accompanied by two republicans, which again holds for agents.

Furthermore, we can ask how panels of purely ideological or party-less judges  would look, by making $\alpha_L \gg \alpha_P \ge 0$, and at purely non ideological judges or law-less, by taking   $0\le \alpha_L \ll \alpha_P$. The results, shown in figure \ref{bandeiraspanel2juc}, show that judges are neither totally ignorant of the law nor free from ideological bias. 

With this tool, we obtain a similarity between the empirical signatures and the 
model signatures by making the choices that the agents are (i') quite courteous towards those of another party and certain about this trust, 
(ii') that there must be a sizable contribution of the common vector $\bm{L}$ to the initial conditions, (iii') that there
is also a sizable contribution of the party bias $\bm{P}$.

A better numerical fit could be done by changing the values of the $\alpha$ parameters, but we restrain  from trying to read into the model a more
realistic replication of the system than what should at this point be considered reasonable. Quantitative evaluation of the components of the Law and the Party can be done for individual courts and is currently under study.

\begin{table}[htb]
\centering
\begin{tabular}{l p{5cm} p{5cm} p{5cm}}
        & Hypothesis                & Example                                   & Interpretation \\
  \hline
  (i)   & ideological voting        & $\theta_{Aaa,Bbb}$ is the largest angle   & Largest differences between $Rrr$ and $Ddd$.\\
  
  (ii)  & ideological dampening     & $\theta_{Baa,Aaa}<\theta_{Abb,Aaa}$       & $D$ in $Drr$ is more conservative than $R$ in $Rdd$.\\
   
        &                           & $\theta_{Abb,Bbb}<\theta_{Baa,Bbb}$       & $R$ in $Rdd$ is more liberal than $D$ in $Drr$.\\
  
  (iii) & ideological amplification & $\theta_{Bbb,Bab}<\theta_{Bbb,Baa}$       & $D$ in $Drd$ is more liberal than $D$ in $Drr$.\\
   
        &                           & $\theta_{Aaa,Aab}<\theta_{Aaa,Abb}$       & $R$ in $Rrd$ is more conservative than $R$ in $Rdd$. \\
\end{tabular}\caption{Some of the evidence for the three hypotheses in the Courteous and Certain scenario.   See 
figure \ref{bandeiraspanel1juc} for the simulation and  for empirical data obtained from  \cite{Sunstein2006}.
}
\label{tabelaJuiz}
\end{table}

\section{A society of $N$ agents\label{Nagents}}

\begin{figure}[ht]
\includegraphics[width=\linewidth]{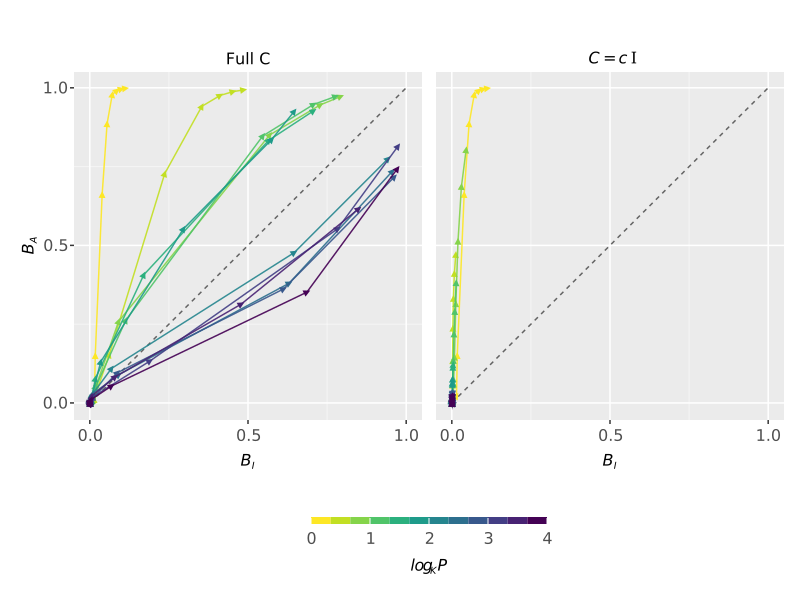}
\caption{ $B_A$ versus $B_I$ for different sizes of the agenda,  as a function of time $\tau$ measured in units of the number of degrees of freedom of the society $N(N+K-1)$ . On the left, agents use the full covariance (tensorial update). For small agendas the affective polarization sets in before the ideological polarization. This is reversed for larger agendas. Right: agents use the simple covariance proportional to the identity matrix ($\bm C= c\bm I$, vector update). For small agendas the system lingers in a disordered ideological state without polarization, but reestablishes affective balance and therefore polarization. For large agendas the system remains in the glassy unbalanced situation in both sectors.}
\label{balance-flow-matrixcov}
\end{figure}

Our next topic is a society of $N$ agents, interacting on the full graph, exchanging $\pm 1$ opinions about $P$   issues $ \bm x$ in $K$ dimensions, with components independently and randomly chosen with standard Gaussian distributions and then normalizing to $|\bm x|=1$. 
As in the case of two agents a variety of behaviors may occur and by no effort was done to exhaust the possible interesting cases. 

We are interested in the polarization of the society into factions and see that two possible types occur, grouping  by affinity of ideology or by affinity of affection. 
These are also discussed in the literature of political science \cite{KlarAffective} \cite{IyengarAffective} 
and  are related to the problem of current interest: Do people disagree because they hate each other or hate each other because they disagree? Interestingly, the answer is it depends.  The dynamical process for each sector can have different characteristic times depending on the ``complexity'' of the agenda under discussion measured in a simple manner by $P$.

In addition to polarization we probe into the breakdown or establishment of balanced norms of trust  \cite{Heider}  \cite{HarariBalance} and balance of opinions in triads of agents. With the same definitions as before for  $\varepsilon_{j|i} = \Phi(h_\mu)$, define $\upsilon_{j|i} = 1-2\varepsilon_{j|i}$, which is positive/negative for trust/distrust. For any three agents \cite{Heider},  the trust relation is balanced 
if  $b_{ijk} := \upsilon_{j|i}\upsilon_{k|j}\upsilon_{i|k} > 0$. This is also studied in physics of disorder materials and receives the name frustration if negative. Since these are directed graphs and the odd permutation $b_{ikj}$ may be of opposite sign,  consider an odd permutation of indices as a different triplet. Then the affective balance, the average value over the population
\ba
B_A &=& \frac{1}{2N_T}\sum_{\langle ijk\rangle}(b_{ijk}+b_{ikj}),
\ea 
informs the state of affective frustration of a society. $N_T$ is the number of triplets.
A second type of frustration can be studied, related to the opinion alignment measured by the symmetric overlap $\rho_{ij}$. Opinion balance, which  characterizes the state of the society is defined by the average over the population
\ba
B_I &=& \frac{1}{N_T}\sum_{\langle ijk\rangle}\rho_{ij}\rho_{jk}\rho_{ki}.
\ea 
An ordered society (approximate consensus) or a polarized society will have a $B_I$ and $B_A$ close to one but these parameters are close to zero for a highly frustrated society. These are seen in figure \ref{balance-flow-matrixcov} where we show the time progression of the balances for two different societies, according to the type of learning algorithm the agents use. For agents using the complete EDNNA algorithm, with a full covariance $\bm C$, the result depends on  the complexity of the agenda. For a simple agenda the society polarizes rapidly into two factions in the affective sector and then the affective polarization drives the ideological polarization. These agents ''hate'' each other before they disagree on the issues. But with time they come to believe different things, also polarizing in the opinion sector, for no other reason that they learn from whom they like and have reversed learning from those they don't. For more complex  agendas the ideological polarization sets in before and then drives the affinity polarization. This is seen in the dark lines, which run below the diagonal in figure \ref{balance-flow-matrixcov}-Left.
 The time to achieve a balanced society increases as $P$ increases. Times $\tau$ measures the  number of learning interactions per degrees of freedom of the system ($\varphi = N(N+K-1)$) so that we can compare simulations with different parameters. 
 
 On the right side of figure \ref{balance-flow-matrixcov} we show the same graphs but now  the agents use  the simpler algorithm obtained with the covariance $\bm C = c\bm I$, so the geometric Hebbian term is not distorted, but simply scaled by an overall uncertainty scalar $c$. 
This dependence on the agenda disappears when the agents use this simpler algorithm. For them affinity polarization always seem  to occur first and  with slow  behavior annealing form the glassy state to the  ideological polarized state. 

\section{Discussion and Conclusions}
There is an understandable appeal for the use of simple models of agents and  the intuitive collective properties that their societies show. However there is a tacit understanding that psychological elements should inform any attempt to increase an agent's realism. Emotions have to enter the mathematical models. When studying the learning models here exposed, we were not trying to include any type of emotions into the final results. Statistical Physics doesn't deal with emotions. However, the modern formulation of Statistical Physics, shows it to be a framework for analyzing systems that attributes probabilities to the different states in a manner that tries to be as ignorant as possible while respecting experimental or informational constraints, thus refraining from incorporating unwarranted evidence. It is thus cast as an example of Information theory. It is not about the physical reality itself of a system but what we can infer about it,  based on the available information. The analysis of the learning algorithm, although in a simple way, finds variables that have a flavor of  proto-emotions,  primitive precursors of what might be called emotions. We can probe the agent and identify that if a dissonance occurs, it elicits changes in at least one of the sectors.  Brains evolved under a myriad environmental pressures, some of them related to efficient information processing. Our agents are optimized by adhering to entropic dynamics and thus process information in their simple world in an optimal way. It is plausible to roughly describe the parallel as metaphoric or to ascribe it to general principles of information processing.  It allows to speculate on some of the reasons of the origin of emotions from an evolutionary point of view: without emotion modulation, learning is deficient. This is totally different from explaining why we feel a surprise, which is is not our aim, which needs a totally different set of theoretical tools that we don't have. We clearly remained tied to a third person perspective of the description of emotions.

The entropic dynamics for neural networks architectures, EDNNA analysis provides one way to obtain such optimized  learning algorithm. But they can be obtained with much more labor as the result of an evolutionary process, similar to that shown in \cite{Neirotti2003}, which deals only with the opinion sector. Of course we have dealt with a small group scenario, that of rule to be  learned  and followed by members of this group,  leading to the opinion sector and a mechanism to emit opinions about issues.  Allowing for cheating by the emitters results in a defense mechanism that appears in the assignment of a level of distrust to other members of the group.  Once the two agent interaction is  defined, we construct a society. 

Since Bayesian probability is simply the name of probability theory from an information point of view,  we stress that this not {\it the} Bayesian algorithm. There is no such thing. Different information scenarios lead to different Bayesian algorithms. Failure to understand this may suggest that humans can't be described by Bayesian modeling. The problem in modeling humans is not that information theory does not apply but that the informational structure is poorly determined in the model. 
The  simpler covariance model results  in figure \ref{balance-flow-matrixcov}Right, where the covariance is just a multiple of the identity, leads to algorithms that are still just as Bayesian, but under a different set of constraints.

Statistical description of the behavior of the judiciary agents, such as \cite{Sunstein2006}  holds the promise of identifying extreme and maybe intolerable political biases. We have restrained from trying to make precise estimates of the amount of political bias or the contribution of the law, trying only to present a strategy to future quantification of the individual courts. It remains clear that without some amount of political bias and without some amount of common law, the agents behave very differently from human judges. It may come as a welcome surprise that the model indicates that only agents that trust agents from not only their party, but the opposing one, and in addition are quite certain of that, have some similarity to the statistical behavior of the human judges. 

Questionnaires are a central method in the humanities to gather information and these neural network agents are capable to include   such scenarios \cite{CaVi2011}.

Lincoln said  \cite{Lincoln} in 1848, ``The process is this: Three, four or half a dozen questions are prominent at a given time, the party selects its candidate, and he takes his position on each of these questions.'' Then opposing parties can form around polarized simple ideological agendas, in his case, the prominent issue under discussion was  whether the new states would be free from slavery. For the agents, affective affinities establish quickly and  proceed to push on a slower time scale to shared opinions and its consequent polarization around ideologies. As the agenda becomes more complex, larger $P$, there is a reversal. The ideological  polarization occurs first,  driving the slower  affective polarization.  The reduction of  times to polarize that accompanies the  reduction of the agenda is in accordance with examples discussed in \cite{FiorinaPolarization} where rapid partisan polarization is not accompanied by rapid changes in positions on economic policy issues. But this doesn't happen for the simpler scalar covariance agents. Affective polarization sets in first and ideological polarization comes later at a much slower scale.

  We have used distrust as a portmanteau for several associated concepts and a refined analysis  is beyond our capacity, see \cite{YamagishiTrust} \cite{InterdisciplinaryPerspectivesonTrustSchockleyEdit} for a extended treatment of these subjects. 
  
  The neural networks we use are certainly simple and limited, despite the rich learning algorithms they use. The EDNNA formalism applies to any architecture, however it becomes intractable fast with architecture complexity.  We believe that agents with more complex architectures and their appropriate EDNNA learning algorithms,  and a larger set of microscopic variables, will display more characteristics that could pose as psychological traits. While this might be true, it would be difficult to identify them and even harder to interpret them without the acquired experience from the simple case architecture. 
  To conclude we stress that proto-emotion markers appear as necessary for efficient learning. 

{\bf Acknowledgment:} We thank A. Caticha, O. Kinouchi, R. Vicente, M. Copelli, JP Neirotti  for discussions. We thank A Martins for pointing out references about ABMs during  FA's PhD defense. FA received financial support from a  Conselho Nacional de Desenvolvimento Científico e Tecnológico (CNPq) PhD fellowship. This work recieved partial support from CNAIPS-NAP USP.

\bibliographystyle{unsrtnat}
\bibliography{EDNNA_frust}  

\end{document}